\documentclass[letterpaper,conference]{IEEEtran}

\ifCLASSINFOpdf
   \usepackage[pdftex]{graphicx}
   \DeclareGraphicsExtensions{.pdf,.jpeg,.png}
\else
\fi
\usepackage{algorithm}
\usepackage{cite}
\usepackage{amsmath,amssymb,amsfonts}
\usepackage{graphicx}
\usepackage{textcomp}
\usepackage{algpseudocode}
\usepackage{siunitx}
\usepackage{subfig}
\usepackage{amsmath}
\usepackage[compact]{titlesec}
\usepackage{cite}
\usepackage[draft]{changes}
\usepackage{booktabs}
\usepackage{textcomp}
\usepackage{balance} 

\usepackage{blindtext}
\usepackage{wrapfig}
\usepackage{amsmath}
\newcolumntype{Z}[1]{>{\centering\arraybackslash\hspace{0pt}}m{#1}}

\usepackage{array}
\usepackage{ragged2e}

\usepackage{pbox}

\usepackage[draft]{changes}

\begin{document}
\linespread{1}
%
% paper title
% Titles are generally capitalized except for words such as a, an, and, as,
% at, but, by, for, in, nor, of, on, or, the, to and up, which are usually
% not capitalized unless they are the first or last word of the title.
% Linebreaks \\ can be used within to get better formatting as desired.
% Do not put math or special symbols in the title.
\title{Cryptocurrency Solutions to Enable Micro-payments in Consumer IoT}

\author{\IEEEauthorblockN{Suat Mercan\IEEEauthorrefmark{1},
Ahmet Kurt\IEEEauthorrefmark{1}, Enes Erdin\IEEEauthorrefmark{2}, and Kemal Akkaya\IEEEauthorrefmark{1}} 
\IEEEauthorblockA{\IEEEauthorrefmark{1}Dept. of Elec. and Comp. Engineering, Florida International University, Miami, FL 33174\\ Email: \{smercan,akurt005,kakkaya\}@fiu.edu}
\IEEEauthorblockA{\IEEEauthorrefmark{2}Department of Computer Science, University of Central Arkansas, Conway, AR 72035  \\Email: eerdin@uca.edu}
%\IEEEauthorblockA{\IEEEauthorrefmark{3}Dept. of Elec. and Comp. Engineering, Florida International University, Miami, FL 33174, Email: mncebe@gmail.com} 
%\IEEEauthorblockA{\IEEEauthorrefmark{4}Dept. of Elec. and Comp. Engineering, Florida International University, Miami, FL 33174, Email: kakkaya@fiu.edu} 
}

% make the title area
\maketitle

% As a general rule, do not put math, special symbols or citations
% in the abstract
\begin{abstract}

The successful amalgamation of cryptocurrency and consumer Internet of Things (IoT) devices can pave the way for novel applications in machine-to-machine economy. However, the lack of scalability and heavy resource requirements of initial blockchain designs hinders the integration as they prioritized decentralization and security. Numerous solutions have been proposed since the emergence of Bitcoin to achieve this goal. However, none of them seem to dominate and thus it is unclear how consumer devices will be adapting these approaches. Therefore, in this paper, we critically review the existing integration approaches and cryptocurrency designs that strive to enable micro-payments among consumer devices. We identify and discuss solutions under three main categories; \textit{direct integration}, \textit{payment channel network} and \textit{ new cryptocurrency design}. The first approach utilizes a full node to interact with the payment system. Offline channel payment is suggested as a second layer solution to solve the scalability issue and enable instant payment with low fee. New designs converge to semi-centralized scheme and focus on lightweight consensus protocol that does not require high computation power which might mean loosening the initial design choices in favor of scalability. We evaluate the pros and cons of each of these approaches and then point out future research challenges. Our goal is to help researchers and practitioners to better focus their efforts to facilitate micro-payment adoptions.

\end{abstract}

\maketitle

\section{Introduction}

Internet of Things (IoT) from tiny sensors to autonomous cars are becoming an indispensable part of life. To create a buoyant and effective IoT ecosystem, it is significant to enable data and service sharing which, in turn, will require making device-to-device payments to the provider for the services such as parking, vehicle charging, sensor data sale, internet sharing, vending machine among others \cite{huckle2016internet}. Since this ecosystem will be dependent on financial micro-transactions among digital objects/devices, a \textit{reliable} payment system without \textit{human intervention} is desirable for a seamless experience. As IoT evolves over time, cryptocurrency can play a critical role by serving as digital money in creating such an ecosystem. The successful integration of IoT and cryptocurrency technologies will foster the revolution by introducing novel consumer applications %in daily life and machine-to-machine economy 
such as enhanced shopping experience for consumers, automated payment among sensing devices, programmable financial transactions for electric vehicles or drones, etc. Nevertheless, this brings many new challenges to be tackled %as it is imperative to have \textit{scalable, cost-efficient} and \textit{real-time} cryptocurrency solutions 
to create an environment that buyers and sellers can perform frictionless transactions in an IoT environment.

%Distributed Ledger based cryptocurrencies ensure a secure payment model as they prevent frauds and double spending since the transactions are secured using cryptographic techniques, and distributed ledger provides non-repudiation in case of conflicts. They can handle transactions in a reasonable time and cost in comparison to today's international money transfer, however, they, including mainstream cryptocurrencies such as Bitcoin and Ethereum, suffer from lack of wide adoption \cite{croman2016scaling} due to its impracticality in daily micro-payments. 

Most of these challenges stem from the way the current distributed ledger based cryptocurrencies are designed \cite{croman2016scaling}. Their design is not feasible to create a machine-to-machine economy for enabling micro-payments due to a number of reasons: 1) \textit{Scalability} as they have limited performance in number of transactions they can handle in a second. For instance, the theoretical maximum throughtput in Bitcoin is calculated to be 7 transactions per second  which is far lower than what Visa or MasterCard can process; 2) \textit{High transaction fees}, which is not attractive for micropayments, and; 3) \textit{Long block confirmation times} as it takes 10 minutes to approve a block of transactions for Bitcoin. In addition, there are other major challenges due to the limitations of the IoT devices. %the integration with IoT devices is challenging because of the computation, communication, and storage limitations. 
Despite the heterogeneity, IoT devices are characterized as small lightweight devices, they are mostly resource constrained and most of them are not be capable of running a full node (e.g., 250 GB storage is required for running a full Bitcoin node). Robust Internet connection and  high computation power are also required to receive and verify new blocks. Therefore, lightweight solutions are needed to enable resource-constrained IoT devices to utilize cryptocurrencies. To this end, recent years witnessed several efforts from the research community to offer solutions to this emerging problem. However, there seems to be no convergence among these solutions to offer a clear path for the benefit of consumers. 

\begin{figure*}[htb]
\vspace{-3mm}
    \centering
    \includegraphics[scale=0.8]{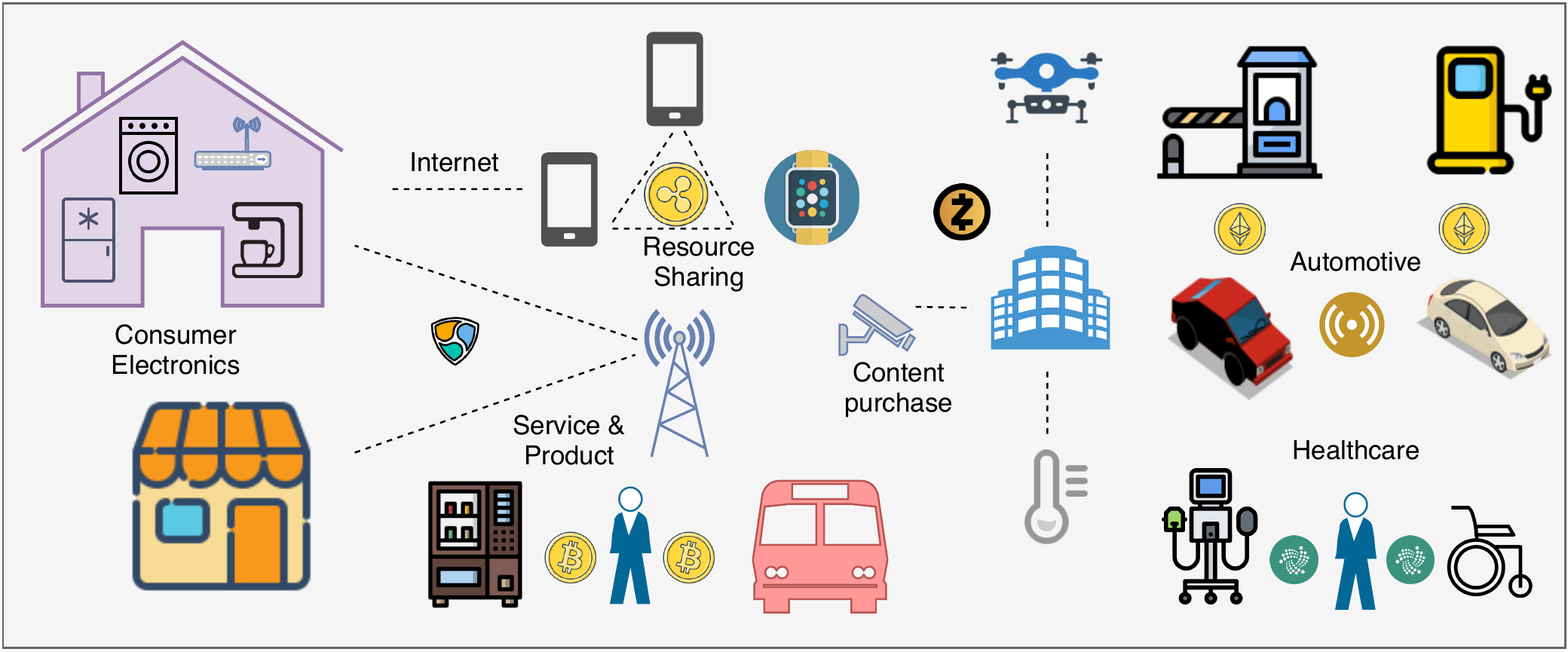}
    \caption{Machine-to-Machine Economy.}
    \vspace{-3mm}
    \label{fig:m2m}
    \vspace{-2mm}
\end{figure*}

Therefore, in this paper, we critically review the solutions that aim to integrate IoT with existing cryptocurrencies and various designs that claim to be IoT friendly. Our goal is to come up with a guideline to be able to assess the benefits and impact of existing solutions. We basically classify the approaches that we find promising and useful to remedy some of the aforementioned problems into three categories: The first category applies \textit{direct integration} through a trusted gateway in order to connect IoT devices to the cryptocurrency network. This does not solve blockchain's inheriting problems such as scalability but it helps to connect resource constrained IoT devices to the network. Second category introduced \textit{Payment Channel Networks (PCN)}, known as a second layer solution, which address specifically the scalability issue by allowing infinite number of off-chain instant transactions among the nodes with minimal fee. Finally, under the third category, new designs are explored among which \textit{Directed Acyclic Graph (DAG)} has emerged as an alternative structure to the blockchain. It still holds the idea of distributed structure in a different manner which is a web of confirmation, tangle, instead of a single blockchain. After we provide a summary of these solutions, we make an evaluation of them considering some qualitative metrics. We discuss which of them could be more appropriate depending on the application. Finally, we provide some future research challenges that need to be tackled for successful deployment of these approaches.

The remainder of this paper is organized as follows: Section II gives preliminaries. In Section III, we present direct integration approaches while Section IV explains Payment Channel Network related works. Section V explores various cryptocurrency proposals traded in the market and Section VI presents the evaluation. Section VII concludes the paper.

\section{Preliminaries}

\noindent \textbf{Basics and Applications:} IoT are utilized in many domains to sense and collect data. In the last few years, they become standard consumer electronics like TVs to be used in homes and cities for smart sensing and control. This created use-cases where these devices would offer services that can be consumed by other devices in return for cash. For instance, consumer electronics in a futuristic smart home may shop independently (e.g., a washing machine can order detergent automatically, the fridge may request refill for milk etc.) An electric autonomous vehicle may pay tolls and also sell its LIDAR data to nearby vehicles. A smart water machine at home can order refills. A drone may sell its video and people may be incentivized through their smart phones to share their resources, %\textit{shared things}, 
such as computation power, data, charge, internet which may require receiving payments from the providers. As a real use case, RightMesh \cite{rightmesh} is using \textit{microRaiden} cryptocurrency to create a mesh network in which participants pay to others. Most of the smart city applications such as electric vehicle charging, sensor data selling will also benefit from this integration. All these examples point out to a future where we will witness machine-to-machine (M2M) economy as depicted in Fig.~\ref{fig:m2m}. In such an economy, the main feature will be \textit{decentralization} of payments to have a self managed system without relying or trusting on third parties and dealing with their management. In this sense, \textit{cryptocurrencies} offer great potential as they rely on distributed ledger technologies providing decentralized management of cash without trusting any third parties.
%Cryptocurrencies on the other hand are based on blockchain technology to provide decentralized management of cash without trusting any third parties. 
Starting with Bitcoin in 2009, many new cryptocurrencies are currently in use that deploy variations of blockchain technologies. A successful integration of IoT and cryptocurrency will enable and boost various innovative applications, which in turn bring convenience and efficiency in our lives. However, this raises several challenges as detailed next.

%This type of interactions require a direct machine to machine communication as opposed to current centralized approaches. Although Blockchain based cryptocurrencies is promising in this aspect, existing challenges must be addressed before they are put into practice.

\vspace{2mm}
\noindent \textbf{Desirable Features for IoT Cryptocurrencies:} %\textit{Security}, \textit{decentralization}, and \textit{scalability} are three goals that blockchain designers are aiming to achieve. It is hard to address three of them together as there is mostly a trade-off among these in existing designs. Early blockchain designs, Bitcoin and Ethereum, prioritized \textit{security}, prevention of double spending, resistance to attacks, by using proof-of-work and distributed structure. Although these are the main purposes of a blockchain system, it unfortunately limits the scalability. 
An IoT ecosystem consisting of many devices interacting with each other generates tremendous number of transactions, and thus requires a high throughput. However, most of the cryptocurrency systems including Bitcoin and 
Ethereum are not \textit{scalable}. Consequently, limited number of transactions and high demand increase the \textit{fee} which could be much greater than actual amount being sent in a micro-payment scenario. Ideally, this should be reasonable, close to zero, for the success of a M2M economy. The time required for the approval of a transaction (i.e., \textit{speed},) is another desirable feature, which should be real-time for a frictionless implementation. Finally, the solutions should be \textit{lightweight} in terms of computation and storage requirement to be able to run on an IoT device as these devices are mostly resource-constrained. 

\begin{figure}[htb]
\vspace{-1mm}
    \centering
    \includegraphics[scale=0.6]{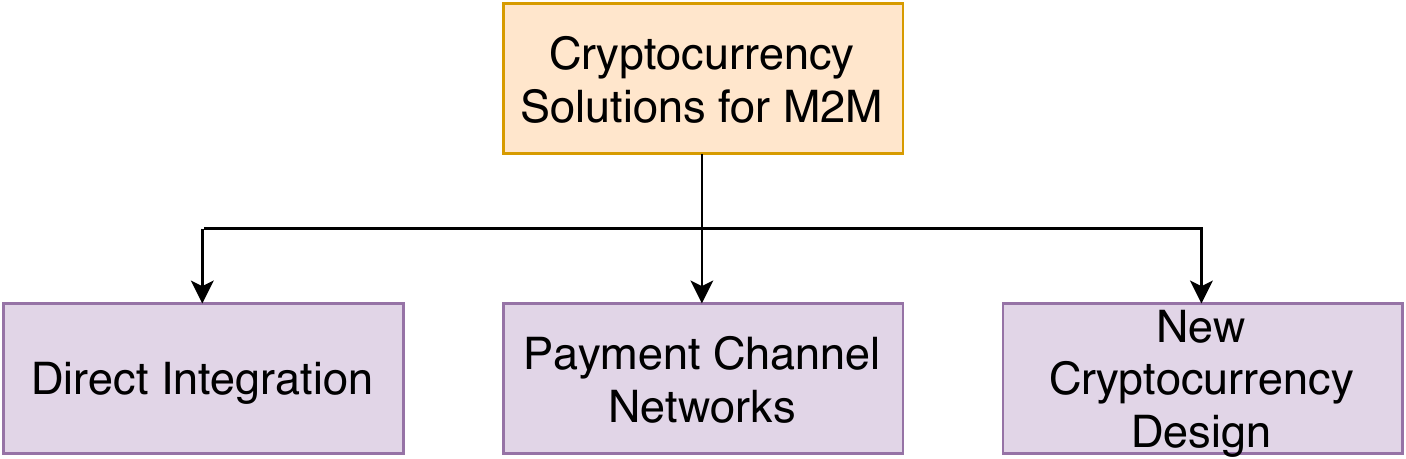}
    \caption{Solution Classifications.}
    \vspace{-1mm}
    \label{fig:sols}
    \vspace{-2mm}
\end{figure}

\vspace{2mm}
\noindent \textbf{Solution Classification:} %As we have not have a perfect solution yet that solves the problems mentioned above, there are various attempts to enable a successful IoT + blockchain integration. Regarding the scalability, most solutions are focusing alternative consensus protocols as the ones relying on participants' computation power (PoW) are limiting and not suitable for IoT. Proof-of-Stake (PoS), Proof-of-Authority (PoA) and Practical Byzantine Fault Tolerance (PBFT) are some popular techniques. However, they are regarded less secure and expand the attack vector. Delegation based voting mechanisms are utilized, Delegated Proof-of-Stake (DPoS), in order to reach the agreement faster, reduce the communication overhead, and release the IoT from doing computation. However, they impose the centralization concern. 
We divide the efforts into three broad classes to address the above issues: 1) Direct Integration; 2) PCNs and 3) IoT Cryptocurrencies (Fig. \ref{fig:sols}) as detailed next.

%Light clients are another effective method in reducing node size, they do not get involved with approval process, however they can verify the transactions using Simplified Payment Verification (SPV); and 3) Moreover, single blockchain which produces only one next block can not satisfy the high number of transactions generated, thus DAG-based structures that can approve transactions in parallel have been proposed. Layered organizations, considering heterogeneity of IoT and varying requirement of each application domain, consists of sub-blockchains aiming to enable cross-chain operations.

%Thus, there is a need for lightweight method to reach agreement on the status of transactions. 
 %Solutions that can approve transactions in parallel might better fit to such an ecosystem.
%This might loose the security features and expand the attack vector.

\section{Direct Integration to Existing Coins}
Integration of IoT to existing major coins such as Bitcoin and Ethereum is achieved either using a gateway or light client:

\subsection{Gateway-based Integration}
\begin{figure}[htb]
\vspace{-1mm}
    \centering
    \includegraphics[scale=0.7]{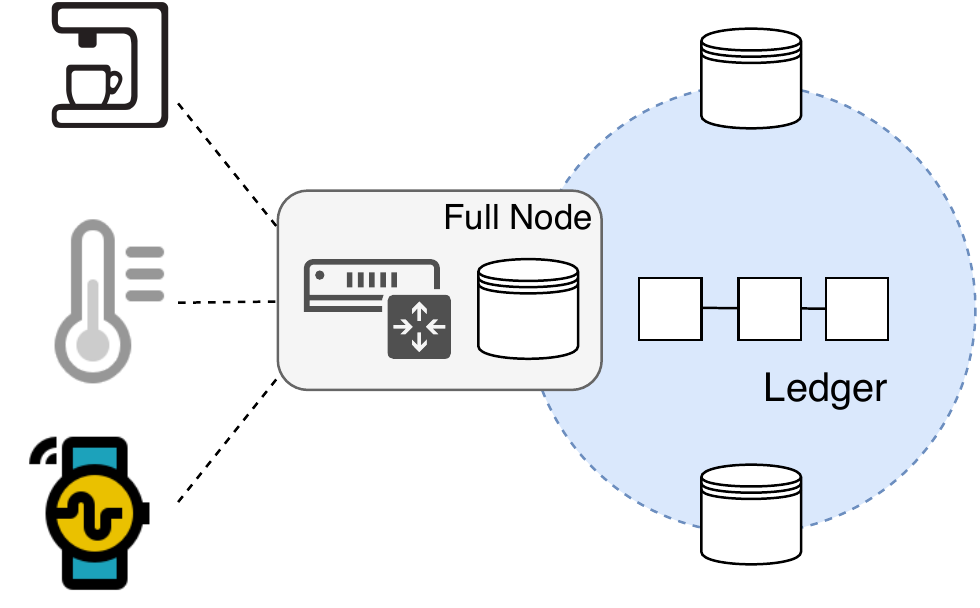}
    \caption{A trusted node is utilized to connect to blockchain.}
    \vspace{-3mm}
    \label{fig:gateway}
  %  \vspace{-1mm}
\end{figure}

In this integration scheme, the IoT device does not run a blockchain node, but relies on another node as shown in Fig. \ref{fig:gateway} which should be operated by the same owner or a trusted third party. IoT devices are registered to the gateway which issues transactions on behalf of them. The communication between IoT and the gateway can be achieved via any protocol such as Wi-Fi, Bluetooth, LoRa etc. For instance, Bitcoin client API (BCCAPI) is designed for such a purpose. The server holds only public key of the client and tracks the client's wallet balance. However, it requires client's consent to make a transaction. Ozyilmaz et al. \cite{ozyilmaz2019designing} exemplifies this concept with Ethereum and LPWAN. Gateway based integration method is utilized by some popular ledgers such as Hyperledger to enable IoT devices to input their data into the system through a peer. %Depending on the trust level of the gateway, various schemes have been developed. 
The communication between the device and the gateway must be secured using cryptographic techniques. Any corruption that may happen during transmission or on the device will not be detected by the blockchain.

\subsection{Light Clients}
Another method that has been proposed for IoT devices to communicate with heavy weight blockchains is by using a \textit{light client}. While full nodes store the whole blockchain to verify the integrity of the blocks, these clients do not need to so. For example in Bitcoin, they can still verify the accuracy of a transaction using a method called \textit{Simplified Payment Verification (SPV)} by using only the headers of the blocks which are requested from an untrusted full node. A block header has the hash of all the transactions in that block and very small in size compared to the real block. Using the Merkle tree structures, it is checked whether the transaction is included in the block header or not. Some of the popular light clients are Electrum, BitPay and Geth.

% Even though the storage and computation requirement is less compared to a full node hosting, this approach still requires some computation for verification. Moreover, light client needs to be synchronized with main blockchain which may consume a substantial data especially after connection intermittance. Thus, this approach may only be feasible for IoT devices with enough resources. 

While light clients can help mainstream adoption, they still relies on the existence of other full nodes. They might also suffer from security and privacy issues. For instance, BIP37 protocol, being used to create lightclients, was exploited to make DDoS attacks on full nodes. Even further, Electrum light client was hacked and resulted in financial loss of the users \cite{electrumhack}. To fix these issues, a new light client protocol, namely BIP157/158 was proposed. In this scheme, in addition the the block headers, block filters are used which helps anonymize the requesting client from the full node.
% Electrum \cite{electrum} relies on dedicated servers that index the blockchain. Clients can query the servers to check unspent transaction outputs, however since the server learns about the transaction details, it degrades privacy.

%by allowing light clients to obtain compact probabilistic filters of block content from full nodes and download full blocks if the filter matches relevant data (needs rephrasing).

%Direct integration scheme either through a gateway or using lightclient solely does not solve blockchain's existing challenges. However, it facilitates the IoT device to take place in the ecosystem. Although there is no silver bullet to solve all the challenges mentioned previously, this scenario might be viable for some applications. %By using this method, IoT devices may either interact with public blockchains or establish a private network.

%Electrum

\section{Payment Channel Networks}

\begin{figure}[htb]
\vspace{-2mm}
    \centering
    \includegraphics[scale=0.5]{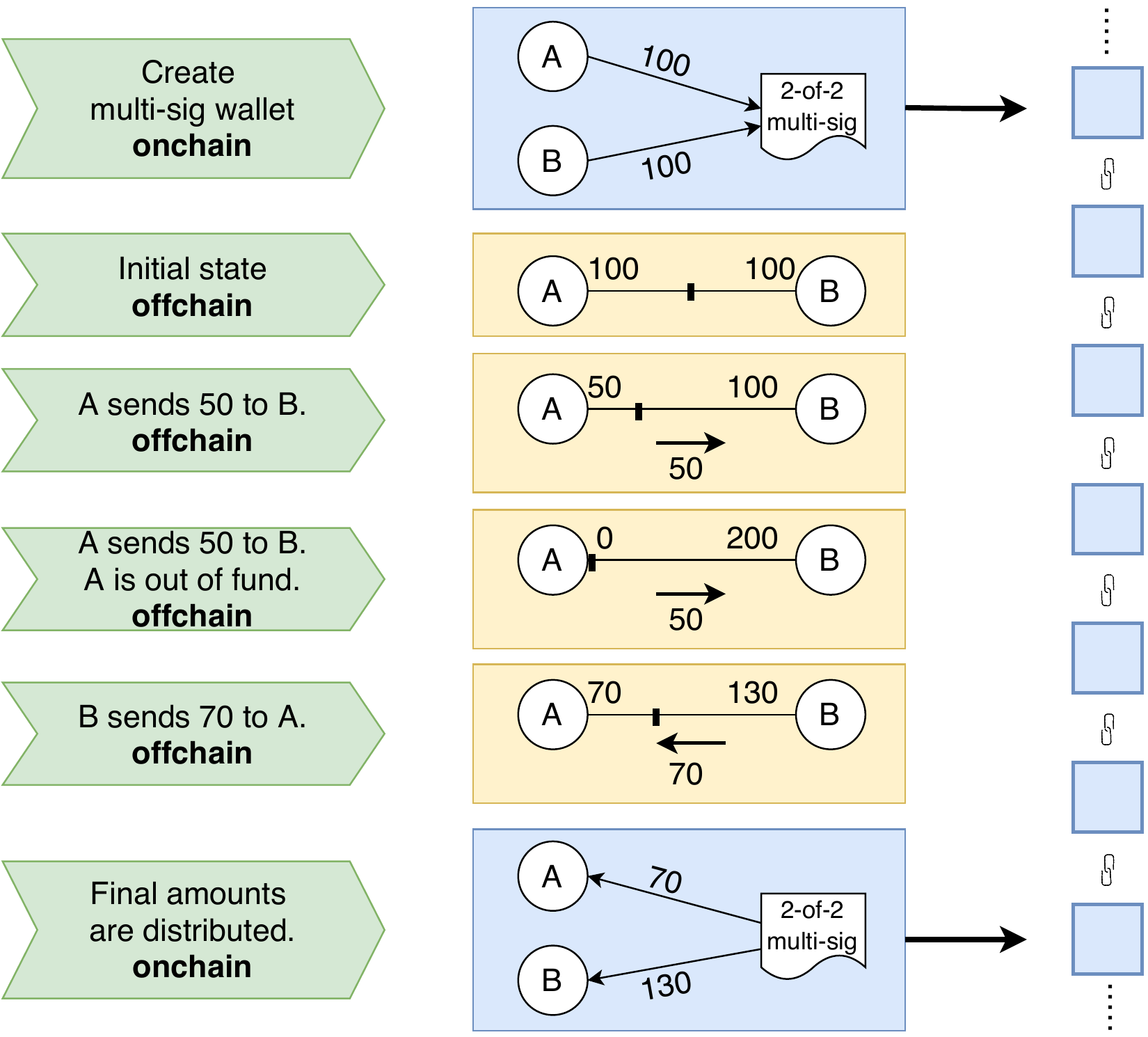}
    \caption{Off-chain transactions are fast.}
    \vspace{-3mm}
    \label{fig:pcn}
    %\vspace{-2mm}
\end{figure}

PCN (also known as off-chain transaction network) concept is a promising solution to solve scalability and latency problems of major blockchains. Thus, it has received great attention from the research community. Lightning Network (LN) \cite{lightningnetwork} for Bitcoin and Raiden for Ethereum are two prominent examples to this concept. The idea is leveraging smart-contracts to avoid broadcasting every transaction to the Blockchain. Instead, the transactions are recorded \textit{off-chain} (Fig. \ref{fig:pcn}) until the accounts are reconciled. Off-chain mechanism brings a huge advantage since the peers do not need to publish every transaction to the blockchain. That is, the payments are theoretically instantaneous. Moreover, as there is no need for frequent on-chain transactions, the transaction fees are protected from fluctuating compared to highly variant on-chain fees. In fact, a transaction fee can be 0 (zero) if the peers agree so. In this aspect, PCN is an attractive solution to use in M2M interactions.

When many nodes come together, the off-chain transaction channels turn into a network of payment channels. Instead of opening a direct channel, a peer makes use of an already established channel to forward money over existing nodes by paying a \textit{transaction fee} as long as a path exists from the payer to the payee. Multi-hop payment scheme also helps users to save on fees when they want to make a payment to someone rather than establishing direct new channel as opening and closing channels incur on-chain Bitcoin transaction fees.

Currently the LN is the most widely used off-chain cryptocurrency payment network which was deployed in 2017 onto Bitcoin mainnet. It has 13,504 nodes and 37,093 channels at the time of writing of this paper. In its current form, LN achieves almost real-time Bitcoin transactions with negligible fees. However, running LN software requires running a Bitcoin client since LN acts as an overlay network on top of Bitcoin. Current Bitcoin blockchain is not suitable for an IoT device to store. Additionally, each newly added block has to be verified by the CPU which will be computationally very expensive for an IoT device. Thus new techniques are required to enable IoT device \& payment channel integration. We investigate some of works on this topic below:

\textit{Ptarmigan} \cite{ptarmigan} is an implementation of the LN protocol specific to the IoT devices.
It is designed and developed with this purpose in mind; to enable \textit{light hardware} to run the full LN protocol natively as well as still complying with the Basis of Lightning Technology (BOLT) specifications. All LN implementations follow the BOLT specifications to enable compatibility between different LN implementations. The mainnet version of Ptarmigan went live in 2019, however, it is still at experimental stage.

\textit{Neutrino} \cite{lightclient} is another Bitcoin \textit{light client} implementation specifically designed for LN. It works based on synchronizing only the block headers and filters instead of the whole blockchain. These filters use Golomb-Rice coding which represent the addresses inside a block and they are much more smaller in size compared to the block headers. %Bitcoin's prior light client protocol BIP37 had flaws such as allowing DDoS vectors on full nodes consequently weakens the security and privacy of the clients. Therefore a new light client protocol, namely BIP157/158 was proposed which fixed these issues by allowing light clients to obtain compact probabilistic filters of block content from full nodes and download full blocks if the filter matches relevant data (needs rephrasing). Thus, neutrino is also a privacy preserving light client. 
Even though some IoT devices might have enough resources to run such light clients, they still have to remain online to synchronize the block headers and filters.

A different approach was proposed in \cite{hannon2019bitcoin} through a protocol that enables IoT devices to open and maintain payment channels with traditional Bitcoin nodes without a view of the blockchain. In this scheme, there are two untrusted third parties which are called \textit{IoT payment gateway} and \textit{watchdog} respectively. IoT payment gateway posts the transactions to the blockchain by creating an additional output. Since there is a chance that IoT payment gateway can post a revoked state to the blockchain, a watchdog is used to inform the IoT device when such thing happens. Both the IoT payment gateway and the watchdog is economically incentivized for their services. %However this system assumes that IoT device can open LN payment channels to the gateway which is the exact problem new research is trying to solve. IoT devices do not have the computational resources to run LN and open payment channels.

Another proposed protocol is ticket-based verification protocol (TBVP) \cite{pouraghily2019lightweight}. % which is proposed to enable IoT devices to participate in financial transaction in an IoT ecosystem. 
In their protocol, the authors introduce two entities which are called \textit{contract manager (CM)} and \textit{transaction verifier (TV)}. Their motivation for introducing these two entities is to separate the blockchain-related operations from blockchain-agnostic operations which are handled by CM and TV respectively. Blockchain-related operations include setting up a smart contract, moving money into the contract, sending commit messages (verified promises) to the contract, closing the contract and claiming the funds. Blockchain-agnostic operations include receiving a commit message from a partner and validating that message. Each IoT device has a CM and TV. 
% In TBVP, there are two separate entities which are called \textit{Gateway} and \textit{Thing}. Main responsibility of the Gateway is to run an instance CM, on the other hand, Things only need to run a TV instance. 
The authors assume every IoT device will work with a separate trusted gateway and these gateways will talk to each other for transactions. %This setup is for cryptocurrency node-to-node transactions and not a typical IoT application setup where an IoT device could connect to any other device for performing the related task.

Different than light clients and protocols, the authors of \cite{robert2020enhanced} proposes a module to integrate LN into an existing IoT ecosystem. In their design, each entity has their own LN node running as part of the system. Once the payment is performed, data provider release the data to the consumer. The proposed system does the integration by decoupling the IoT and LN node and defining the communication API between them. Authors integrated LN into their IoT ecosystem through a payment module that they called \textit{LN module} which consists of a Bitcoin node, an LN node, web-service and web UI. The role of LN module is to create channels between data consumers and owners as well as routing of payments and data. This approach assumes that Bitcoin wallets are hold by the third parties in the IoT marketplace which raises privacy and reliability concerns.

\section{IoT Cryptocurrency Proposals}

In this section, we explore various DLT-based coins, as shown in Table I, traded in the cryptocurrency market and present themselves as IoT compatible solutions. 

\begin{table}[htb]
\vspace{-1mm}
\centering
\caption{IoT Coin Proposals.}
\label{tab:comparative_table}
% \begin{tabular}{|Z{4em}|Z{6em}|Z{4em}|Z{4em}|Z{4em}|} \hline
\begin{tabular}{|c|c|c|c|c|} \hline
 
  ~              & \textbf{Network Type} & \textbf{Structure} & \textbf{Consensus} & \textbf{Additional Info}        \\ \hline \hline
  
 \textbf{IOTA}      & Semi-centralized   & DAG         & PoW  & MCMC            \\ \hline
 \textbf{Nano}       & Semi-centralized & DAG       & ORV    & Nlock-lattice         \\ \hline
 \textbf{IoTeX}       & Semi-centralized & blockchain           & Roll-DPOS  & hierarchical            \\ \hline
 \textbf{ITC}       & Decentralized & blockchain           & PBFT    & Hardware-based          \\ \hline
 \textbf{Walton}       & Semi-centralized  & blockchain           & PoC  & Hardware-based            \\ \hline
\end{tabular}{} 
\vspace{-1mm}
\\
\flushleft
\vspace{-1mm}
\end{table}

\begin{figure}[htb]
\vspace{-3mm}
    \centering
    \includegraphics[scale=0.6]{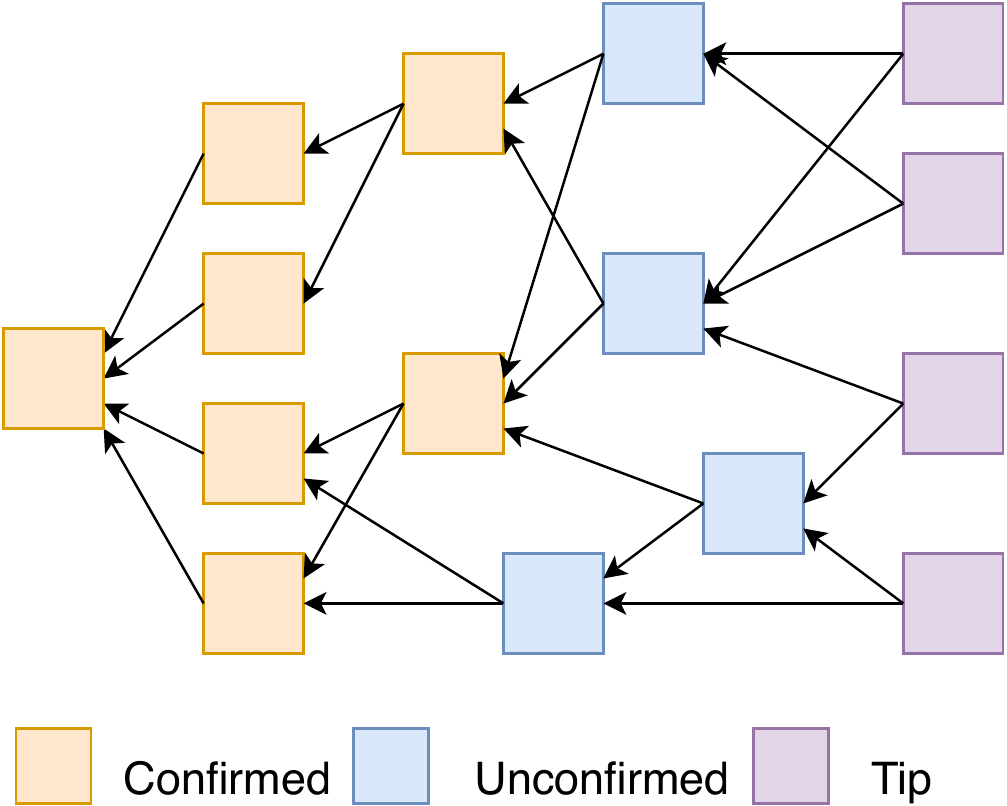}
    \caption{Directed Acyclic Graph for IOTA.}
    \vspace{-3mm}
    \label{fig:dag}
    %\vspace{-1mm}
\end{figure}

\textbf{IOTA}: IOTA \cite{popov2016tangle} is one of the first coins adopted DAG as an alternative to single blockchain structure. %It has recieved attention from community including academia and has the highest marketcap in this category. 
It is considered highly scalable, near-instant and low-fee cryptocurrency with a focus on micro-payments in IoT universe. The idea, \textit{Tangle}, aims to improve the scalability by enabling parallel transaction approval. Block concept which contains multiple transactions as in blockchains does not exist, but each transaction is connected to two previous transaction by the confirmation which creates a web of connections. Each vertex in Fig. \ref{fig:dag} represents a transaction which should validate two previous transactions. Whenever someone wants to add a transaction, s/he should validate two others by selecting according to Markov Chain Monte Carlo (MCMC) algorithm. 
%The node must check whether these transaction are valid or in conflict. 
Thus, transactions issuer are at the same time approvers. 
%Then, he can braodcast his own transaction to the network. 
%The consensus is based on the number of approvals received from other transactions. 
Once a transaction is referred by enough number of transactions, it is considered that the system has reached agreement on this record. That is, consensus is based on cumulative PoW of stacked transactions. %Higher number of approvals means higher level of confidence. 
The system does not have specific miners because users confirms each other, thus the transactions are feeless. The IOTA node still needs to find a nonce for PoW which is much lighter than Bitcoin and thus this makes it scalable and cost efficient. It may take several minutes for a transaction to get approved. The ledger is completely transparent, thus transactions are not private. 
The current definition of IOTA requires that a transaction must be directly or indirectly signed by the coordinator node, then it is assumed 100\% confirmed. %These confirmations are generated as batches known as milestone in every two minutes. 
The existence of coordinator ensures the security of the transactions while the system converges to centralized approach and might be vulnerable to failures \cite{silvano2020iota}. 
%In spite of these, there are many proof-of-concept implementation and investigation on IOTA which shows a potential and interest from researchers. So, it is perceived as a promising solution.

IOTA is not the only coin that utilizes DAG, it has been adopted in various forms. The approach is criticized in general by blockchain supporters as it does not rely on heavy hash-based PoW. However, proponents view this feature as a solution to post-quantum threats \cite{silvano2020iota}.

\textbf{Nano:} This is also known as \textit{RaiBlocks} \cite{nano} adopting DAG approach. It aims to be a high-performance cryptocurrency that can run on low-power hardware. The ledger design is based on individual accounts each of which has their own blockchain. The global ledger consists of these individual chains called \textit{block lattice} in the form of DAG. An account is represented by public portion of a public/private key pair. Transactions are signed by private key of the account holder to confirm that the content is generated by the user. %Each individual account is distributed to the network. 
A transfer from one account to another requires creating two blocks; \textit{send} and \textit{receive}. The former one debits the sender's account as the latter credits the receiver account. Once a sending transactions is confirmed, it can not be reversed. %Even the receiver is offline, it can create the matching transaction to receive the payment. The main block has all the coins in the beginning which has been transferred to other accounts by \textit{send} transaction. 
Nano adopts a consensus mechanism, \textit{Open Representative Voting}, which is delegation based and balance-weighted vote mechanism, a variant of Delegated Proof-of-Stake. Account owners select a delegate to vote on behalf of him and can change it any time. The weight of the voter is proportional to the sum of investment of the users who delegated to him. %The ones who has more investment in the network is given more power because they have higher incentive to maintain the network. The representatives should always be online and vote for the validity of the transaction. 
As the votes of principal representatives are broadcast and counted, a transaction which has enough vote will be counted as confirmed. %A node can be principal representative if he has more than \%0.1 weight in the network. 
Hardware requirements of such a node is 4GB RAM, 200 Mbps bandwidth %(2TB or more of available monthly bandwidth), SSD-based hard drive 
with 80GB+ of free space. So, IoT devices rely on principal nodes in terms of system consistency whose success depends on the existence of enough number of nodes.

\begin{figure}[htb]
\vspace{-3mm}
    \centering
    \includegraphics[scale=0.6]{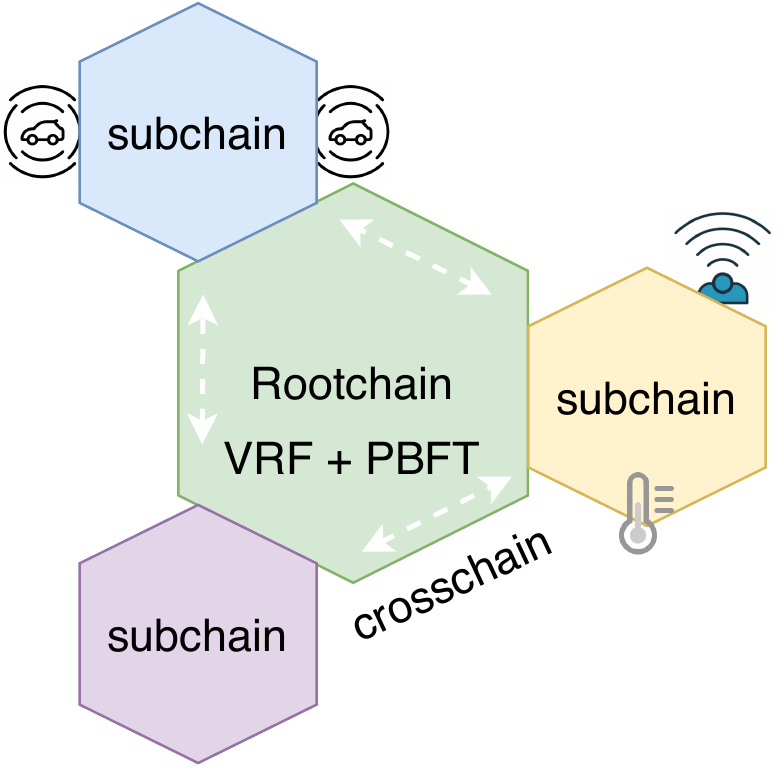}
    \caption{Rootchain and subchains.}
    \vspace{-2mm}
    \label{fig:rootchain}
    %\vspace{-1mm}
\end{figure}

\textbf{IoTeX:} This approach positions itself as IoT-oriented, privacy-centric and scalability maximizing \cite{iotex}. It proposes a model called "seperation of duties" by creating sub-blockchains in order to make it manageable in size. %It defines itself as network of blockchains organized in hierarchy which is similar to the definition of internet as "network of networks". 
Each node interacts only with a specific group and each group may have its own features which optimizes different priorities as IoT devices also differ in capacity. Each blockchain may contact another one through root blockchain as shown in Fig. \ref{fig:rootchain}.  Rootchain is public while sub-chains may be public or private. The insight behind the idea of allowing heterogeneous sub-blockchains is the fact that the requirements of various application domains is different as there is no solution that satisfies all. %For instance, the priority of a smart home is different than a large size IoT deployment as well as capacities of used devices. 
Thus, they may be optimized towards different directions. Moreover, hosting all IoT devices in one blockchain will make it grow fast which might impact the performance. The rootchain has three main functionalities, \textit{relay} of value and data accross subchains,  \textit{supervision} of subchains and \textit{settlement} of payments. The rootchain proposes Randomized Delegated Proof-of-Stake (Roll-DPoS), which is the combination of Verifiable Random Function (VRF), DPoS and Practical Byzantine Fault Tolerance (PBFT). Participants of the network vote to elect candidates which will be in the committee for a period of time. %Currently, 97 candidates are selected, 11 of which randomly chosen using VRF to join the voting process in each iteration. 
VRF is fed with the hash of previous block and node's private key, then others can verify the committee members using their public keys. In each round, committee members propose blocks which are voted based on PBFT and added to the chain if approved. %In order to make easier the verification of blocks for light clients, it also creates periodic checkpoints as done in Ethereum, called \textit{epoch}. 

\textbf{ITC}: IoT Chain (ITC) \cite{iotchain} is another DAG-based currency designed as a lite operating system. It aims at creating a secure communication environment as well as asset transfer. All nodes are considered equal, so decentralized approach is targeted which protects privacy. It adopts PBFT and Simplified Verification Protocol (SPV), which allows to verify transactions by using headers only, to address the size expansion problem. The developers aims designing specific chips embedded with security credentials which means devices joining the system mush have a built-in ITC. 

\textbf{Walton Chain:} This approach \cite{walton} focuses on data reliability with tamper-proof devices. It proposes a hardware-based and hierarchical blockchain. The intuition behind using hardware-based solution is that software solutions are vulnerable to compromises. It aims to ensure data authenticity as Blockchain can guarantee the data integrity after it has been recorded. They have a chip design embedded with security credentials and compatible with various communication protocols. It uses Waltonchain Proof-of-Contribution (WPoC) which is defined as the combination of PoW, PoS and Proof-of-Labor (PoL). First two is used on the parent chain as the PoL is defined to perform coin exchange between child chains. 

% Machine Exchange Coin(MXC) \cite{mxc} is actually creating a platform focusin on inter-chain data market by defining a long range communication protocol (Machine Exchange Protocol) in addition to coin. The sensors are required to connect to a gateway using defined protocol to send data. The gateway is part of the network, holds the wallet, thus handles the payment. It does not propose a new blockchain but allows other permissionless platforms such as Ethereum, Stellar to be used.

% \textbf{Byteball :}
% IBM watson
% Exxor

% Vechain
% https://gobyte.network/

% Power Ledger: It is designed for energy trading. It is not only for payment, but also keeps details of transactions.

\section{Qualitative Evaluation}

In this section, we evaluate and compare the three categories of solutions by using some qualitative metrics as shown in Table II. We picked the following metrics:  \textit{Transaction per second (TPS)} is the total number of transactions that the system can confirm which is the major issue that early designs suffer because of PoW and limited blocksize. PCN has substantially solved this problem as new designs mostly focus on this and made some improvements. \textit{Transaction fee} metric indicates the amount charged to the sender for the transaction. Direct integration based approach does not impact the original fee (charged by Bitcoin), thus it is high, while PCN enables very low fee. New coins are designed mostly to provide feeless transactions by using alternative PoW and eliminating mining. \textit{Confirmation time} refers to the total time to get transaction validated. As seen, PCNs and new coins offer faster speed as they address the slower transaction times of existing coins. \textit{Security} is related to the prevention of double-spending and robustness of chain to various attacks. As Bitcoin and Ethereum are highly secure, implementation of integration protocol will determine the security level. PCN (i.e., LN) has been investigated deeply in last years, and it is perceived a secure design based on HTLC. Newly designed coins have not been able to establish trust yet. As the general approach is converging to semi-centralized structure, they are criticized for being less secure. \textit{Privacy} is defined as the anonymity of sender and receiver information. Bitcoin and Ethereum are not privacy preserving as the transactions are transparent. Moreover the approaches that apply gateway integration may also let the gateway to observe the transactions. PCN provides privacy by using onion type routing. 

% \begin{table}[htb]
% \centering
% \caption{Comparison of Approaches.}
% \label{tab:comparative_table2}
% \begin{tabular}{|Z{6em}|Z{6em}|Z{6em}|Z{6em}|} \hline
% % \begin{tabular}{|c|c|c|c|} \hline
 
%   ~              & \textbf{Direct Integration} & \textbf{PCN-based} & \textbf{New Coins}        \\ \hline \hline
  
%  \textbf{TPS}      & Low & High & High            \\ \hline
%  \textbf{Transaction Fee}       & High & Low & Low                   \\ \hline
%  \textbf{Confirmation Time}       & High & Low & Low           \\ \hline
%  \textbf{Security}       & High & High & Vary           \\ \hline
%  \textbf{Privacy}       & Low & High & Vary           \\ \hline
% %  Integration Scheme       & Third party/ Trusted Gateway & Third party/ Gateway/ Light client & Direct/ Gateway           \\ \hline
% %  Ease of deployment       &  & Yes & No           \\ \hline
%  \textbf{Bitcoin/Ethereum Compatibility}       & Yes & Yes & No           \\ \hline
% \end{tabular}{} 
% \vspace{-1mm}
% \\
% \flushleft
% %\vspace{-2mm}
% \end{table}

\begin{table}[htb]
\centering
\caption{Comparison of Approaches.}
\label{tab:comparative_table2}
\begin{tabular}{|c|c|c|c|} \hline
    ~                 & \textbf{Direct}      & \textbf{PCN}   & \textbf{New}   \\ 
    ~                 & \textbf{Integration} & \textbf{Based} & \textbf{Coins} \\ \hline \hline
\textbf{TPS}          & 7 for Bitcoin        & $\infty$       &  $ > $ 1000    \\ \hline
\textbf{Transaction}  &  $\approx$ \$10      & $ < $ \$0.01   & \$0            \\ 
\textbf{Fee}          &                      &                &                \\ \hline
\textbf{Confirmation} & 10 min               & 1-5 sec        & 1 min          \\
\textbf{Time}         & for Bitcoin          &                &                \\ \hline
\textbf{Security}     & High                 & High           & Vary           \\ \hline
\textbf{Privacy}      & Low                  & High           & Vary           \\ \hline
%  Integration Scheme       & Third party/ Trusted Gateway & Third party/ Gateway/ Light client & Direct/ Gateway           \\ \hline
%  Ease of deployment       &  & Yes & No           \\ \hline
 % \textbf{Bitcoin/Ethereum Compatibility}       & Yes & Yes & No           \\ \hline
\end{tabular}{} 
\vspace{-1mm}
% \\
% \flushleft
%\vspace{-2mm}
\end{table}

When all the solution types are considered, \textit{direct integration} allows to use major cryptocurrencies by inheriting the features of used coins, which means that it will not change their negative sides. Thus, it will not help to create a transaction-intensive M2M economy, but it may still be useful for some scenarios. Although new cryptocurrencies bring promising designs, they have not been tested enough, thus there is no consensus on adopting any of them yet. Among them, IOTA prevails over others as it improves the crucial metrics with its alternative design. However, it still needs to be improved to lower confirmation times. At this stage, PCNs with light client and other integration methods might be the best solution as they offer all the features of new coins in addition to being secure and operable with Bitcoin etc. Although they have not been perfected yet, they promise great potential with their second layer architecture.

\section{Conclusion}

In this paper, we aimed to shed light on IoT and cryptocurrency integration that can foster M2M economy. We reviewed and discussed existing solutions under three broad categories; \textit{direct integration} is utilizing a gateway to communicate with blockchain, \textit{payment channel networks} is an effective method to increase number of transactions and lower the fee and \textit{IoT cyrptocurrency designs} are mainly focusing on novel PoW algorithms that requires less computation resources. We then evaluated these approaches by comparing using various metrics.

The interest in cryptocurrency from formal institutions and governments and wide adoption of IoT towards smartization  will accelerate the efforts for IoT and cryptocurrency amalgamation. In this sense, since the altcoins have not established enough trust, the efforts to integrate Bitcoin and Ethereum to IoT will continue. While the direct integration does not solve the fundamental scalability issue, it can still be feasible for some scenarios. However, novel approaches such as delagated and randomized confirmation models, secure and lightweight consensus algorithms will be investigated in more detail by the community until there is a convergence. The authors of this article considers PCN as a promising solution to particular problem of IoT and cryptocurrency integration as it allows instant and limitless transaction. However, it is not IoT compatible in terms of resource requirement. Thus, we believe that some cryptographic methods such as threshold signature, proxy-encryption should be investigated to achieve a secure integration to Lightning Network.

%cross chain transactions

\bibliographystyle{IEEEtran}

%\bibliography{References.bib}

\end{document}